# Direct vs. Two-Step Approach for Unique Word Generation in UW-OFDM

Alexander Onic, Mario Huemer
Institute of Networked and Embedded Systems, Alpen-Adria-Universität Klagenfurt, Austria
{alexander.onic, mario.huemer}@uni-klu.ac.at

*Abstract*— Unique word OFDM is a novel technique for constructing OFDM symbols, that has many advantages over cyclic prefix OFDM. In this paper we investigate two different approaches for the generation of an OFDM symbol containing a unique word in its time domain representation. The two-step and the direct approach seem very similar at first sight, but actually produce completely different OFDM symbols. Also the overall system's bit error ratio differs significantly for the two approaches. We will prove these propositions analytically, and we will give simulation results for further illustration.

## I. Introduction

CYCLICITY of an OFDM symbol is a necessary condition that needs to be fulfilled in order to be able to perform OFDM transmission in a multipath environment. Traditionally a cyclic prefix (CP) is used to guarantee the cyclicity.

While this method is well examined and understood, there is another possibility to ensure the cyclicity. If a unique word (UW) of length $T_{\mathrm{GI}}$ is chosen in advance and introduced at the end of each OFDM symbol, cyclicity appears, too.

Different to the CP, which is copied and added after IDFT (Inverse Discrete Fourier Transform), the UW is part of the IDFT output, and therefore also part of the DFT interval $T_{\mathrm{DFT}}$. Figure 1 sketches the structure of CP- and UW-OFDM symbols.

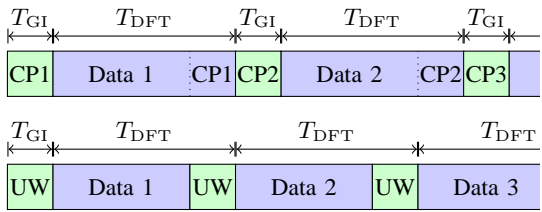

Fig. 1. OFDM symbol structure using cyclic prefixes and unique words

Note again, that the guard interval in figure 1 containing the cyclic prefixes are copies of a part of the payload data and hence random. In contrast the UW is deterministic and known in advance, which allows additional processing.

Summing up, we want to stress some points regarding UW-OFDM:
- Cyclicity is also ensured as in CP-OFDM.
- The UW can be used for synchronization and estimation tasks [1]
- Improved performance regarding bit error ratio compared to CP-OFDM in frequency selective environments [2].
- Almost no loss in bandwidth efficiency [2].

We denote vectors in bold lower case $\mathbf{a}$, frequency domain vectors additionally with a tilde $\tilde{\mathbf{a}}$ and matrices in bold upper case $\mathbf{A}$. The operations $\mathbf{a}^\mathsf{T}$ and $\mathbf{a}^\mathsf{H}$ indicate the matrix transpose resp. conjugate transpose.

In section II we introduce the concepts of UW-OFDM and present two different approaches for UW generation, that come into mind naturally. We will examine the OFDM symbol energies resulting from both approaches in section III and show numerical examples proving these results in section IV. Section V concludes our work.

## II. Unique Word OFDM

The generation of the final OFDM symbol in UW-OFDM differs a lot from CP-OFDM. For one symbol in CP-OFDM the data symbols $\tilde{\mathbf{x}}_d \in \mathbb{C}^{N_d \times 1}$ are loaded onto the subcarriers. Then zero subcarriers are inserted at the band edges and the DC position which we describe by a matrix operation $\tilde{\mathbf{x}} = \mathbf{B}\tilde{\mathbf{x}}_d$ with $\tilde{\mathbf{x}} \in \mathbb{C}^{N \times 1}$ and $\mathbf{B} \in \{0,1\}^{N \times N_d}$. This frequency domain vector is then transformed into time domain via the IDFT operation, which we denote by a matrix operation $\mathbf{x} = \mathbf{F}_N^{-1}\tilde{\mathbf{x}}$ utilizing the $N$-point DFT matrix $\mathbf{F}_N$ with the element of the $m$-th row and the $n$-th column $[\mathbf{F}_N]_{m,n} = \mathrm{e}^{-\mathrm{j}\frac{2\pi}{N}mn}$, where $m,n = 0,1,2,\ldots,N-1$. The guard interval is then formed by copying the last values to the front.

In UW-OFDM the content of the guard interval is known in advance and part of the IDFT operation.

In order to obtain a predefined sequence at the last $N_u$ positions we have to spend at least this amount of freedom on the input side of the IDFT. Thus we define some carriers as *redundant* carriers, which can not be used for data transmission, but have to be loaded with appropriate values $\tilde{\mathbf{x}}_r \in \mathbb{C}^{N_r \times 1}$ to yield the UW at the output.

A transmit symbol can be described by

$$\mathbf{x} = \mathbf{F}_N^{-1} \mathbf{B} \mathbf{P} \begin{bmatrix} \tilde{\mathbf{x}}_d \\ \tilde{\mathbf{x}}_r \end{bmatrix}, \quad (1)$$

where $\mathbf{P}$ is a permutation matrix $\mathbf{P} \in \{0,1\}^{(N_d+N_r)\times(N_d+N_r)}$, that changes the positions of the data and redundant values in an optimum way. Note that $\mathbf{B}$ is now $\mathbf{B} \in \{0,1\}^{N \times (N_d+N_r)}$. As shown in [2] and [3] the choice of $\mathbf{P}$ is crucial for obtaining low energy contributions on the redundant subcarriers, but its particular design is of no further relevance for the investigations in this paper.

Still we haven't had a look at the resulting time domain vector $\mathbf{x}$ and the symbols on the redundant carriers $\tilde{\mathbf{x}}_r$. Here two approaches can be taken into account and will be explained in detail.

*A. Two-step approach*

The two-step approach aims on generating a zero word at the position of the unique word first. In a second step we add the unique word in time domain:

$$\mathbf{x}' = \begin{bmatrix} \mathbf{x}_p \\ \mathbf{0} \end{bmatrix} = \mathbf{F}_N^{-1} \mathbf{B} \mathbf{P} \begin{bmatrix} \tilde{\mathbf{x}}_d \\ \tilde{\mathbf{x}}_r \end{bmatrix} \quad (2)$$

$$\mathbf{x} = \mathbf{x}' + \begin{bmatrix} \mathbf{0} \\ \mathbf{x}_u \end{bmatrix}, \quad (3)$$

with the unique word $\mathbf{x}_u \in \mathbb{C}^{N_u \times 1}$ and the payload $\mathbf{x}_p \in \mathbb{C}^{(N-N_u) \times 1}$.

Splitting up these matrix operations into appropriately sized sub-matrices

$$\mathbf{F}_N^{-1} \mathbf{B} \mathbf{P} = \begin{bmatrix} \mathbf{M}_{11} & \mathbf{M}_{12} \\ \mathbf{M}_{21} & \mathbf{M}_{22} \end{bmatrix}, \quad (4)$$

we can extract the zero word generation and solve for the redundant subcarrier symbols

$$\mathbf{0} = \mathbf{M}_{21} \tilde{\mathbf{x}}_d + \mathbf{M}_{22} \tilde{\mathbf{x}}_r \quad (5)$$

$$\tilde{\mathbf{x}}_r = -\mathbf{M}_{22}^{-1} \mathbf{M}_{21} \tilde{\mathbf{x}}_d$$
$$= \mathbf{T} \tilde{\mathbf{x}}_d \quad (6)$$

by the matrix $\mathbf{T} = -\mathbf{M}_{22}^{-1} \mathbf{M}_{21}$.

Since we define the number of redundant subcarriers as $N_r = N_u$, $\mathbf{M}_{22}$ is quadratic with permuted Vandermonde structure and invertible.

With the result for the redundant subcarrier symbols we can find the following expression for the frequency domain vector $\tilde{\mathbf{x}}$ that is fed into the IDFT:

$$\tilde{\mathbf{x}} = \mathbf{BP} \begin{bmatrix} \tilde{\mathbf{x}}_d \\ \tilde{\mathbf{x}}_r \end{bmatrix} = \mathbf{BP} \begin{bmatrix} \tilde{\mathbf{x}}_d \\ \mathbf{T} \tilde{\mathbf{x}}_d \end{bmatrix}$$
$$= \mathbf{BP} \begin{bmatrix} \mathbf{I} \\ \mathbf{T} \end{bmatrix} \tilde{\mathbf{x}}_d \quad (7)$$

We let $\mathbf{G} = \mathbf{P} \begin{bmatrix} \mathbf{I} \\ \mathbf{T} \end{bmatrix}$ to get the simple expression

$$\tilde{\mathbf{x}} = \mathbf{B} \mathbf{G} \tilde{\mathbf{x}}_d. \quad (8)$$

Following this approach, the final transmit vector is found by computing $\mathbf{BG}\tilde{\mathbf{x}}_d$, changing to time domain by applying the IDFT and finally adding the UW, as initially intended.

Note, that with this approach the UW spectrum is added to the result of (2). So the UW exerts influence on every subcarrier in general, depending on the choice of the UW.

*B. Direct approach*

In contrast to the two-step approach, the unique word can also be generated directly at the output of the IDFT:

$$\begin{bmatrix} \mathbf{x}_p \\ \mathbf{x}_u \end{bmatrix} = \mathbf{F}_N^{-1} \mathbf{B} \mathbf{P} \begin{bmatrix} \tilde{\mathbf{x}}_d \\ \tilde{\mathbf{x}}_r \end{bmatrix}. \quad (9)$$

Following the derivations of the two-step approach we get

$$\mathbf{x}_u = \mathbf{M}_{21} \tilde{\mathbf{x}}_d + \mathbf{M}_{22} \tilde{\mathbf{x}}_r$$
$$\tilde{\mathbf{x}}_r = \mathbf{M}_{22}^{-1} \mathbf{x}_u - \mathbf{M}_{22}^{-1} \mathbf{M}_{21} \tilde{\mathbf{x}}_d$$
$$= \mathbf{M}_{22}^{-1} \mathbf{x}_u + \mathbf{T} \tilde{\mathbf{x}}_d \quad (10)$$

and finally obtain

$$\tilde{\mathbf{x}} = \mathbf{BP} \left( \begin{bmatrix} \mathbf{I} \\ \mathbf{T} \end{bmatrix} \tilde{\mathbf{x}}_d + \begin{bmatrix} \mathbf{0} \\ \mathbf{M}_{22}^{-1} \end{bmatrix} \mathbf{x}_u \right)$$
$$= \mathbf{BG} \tilde{\mathbf{x}}_d + \mathbf{BP} \begin{bmatrix} \mathbf{0} \\ \mathbf{M}_{22}^{-1} \end{bmatrix} \mathbf{x}_u. \quad (11)$$

After transformation of $\tilde{\mathbf{x}}$ into time domain the signal is ready to be sent.

While in the two-step approach the UW is able to influence all subcarriers, here this is not possible. The only impact of the UW in frequency domain can be seen in (10), which is on the redundant carriers, regardless of the actual UW spectrum.

## C. Receiver design

The channel propagation of one OFDM symbol can be modeled with the cyclic channel convolution matrix $\mathbf{H} \in \mathbb{C}^{N \times N}$ and additive white Gaussian noise $\mathbf{n} \in \mathbb{C}^{N \times 1}$ as $\mathbf{r} = \mathbf{H}\mathbf{x} + \mathbf{n}$.

Transforming the received vector into frequency domain and removing the zero carriers using $\mathbf{B}^\mathsf{T}$, we get the disturbed vector of data and redundant subcarrier symbols as

$$\tilde{\mathbf{y}} = \mathbf{B}^\mathsf{T} \mathbf{F}_N \mathbf{r} = \mathbf{B}^\mathsf{T} \mathbf{F}_N \mathbf{H} \mathbf{x} + \mathbf{B}^\mathsf{T} \mathbf{F}_N \mathbf{n}. \quad (12)$$

As presented in [2] an LMMSE estimator can be derived that extracts the data part by

$$\widehat{\tilde{\mathbf{x}}}_d = \widetilde{\mathbf{W}} \widetilde{\mathbf{H}}^{-1} \left( \tilde{\mathbf{y}} - \widetilde{\mathbf{H}} \mathbf{B}^\mathsf{T} \tilde{\mathbf{x}}_u \right) \quad (13)$$

using the spectral influence of the UW $\tilde{\mathbf{x}}_u$, described later in this section, and a Wiener smoother

$$\widetilde{\mathbf{W}} = \mathbf{G}^\mathsf{H} \left( \mathbf{G}\mathbf{G}^\mathsf{H} + \frac{N\sigma_n^2}{\sigma_d^2} \left( \widetilde{\mathbf{H}}^\mathsf{H} \widetilde{\mathbf{H}} \right)^{-1} \right)^{-1}. \quad (14)$$

This suggests the following decoding procedure:
1) Transform received symbol into frequency domain and discard zero carriers.
2) Subtract spectrum influence of the UW, after passing the channel $\widetilde{\mathbf{H}} \mathbf{B}^\mathsf{T} \tilde{\mathbf{x}}_u$.
3) Apply zero-forcing equalization $\widetilde{\mathbf{H}}^{-1}$.
4) Apply Wiener smoothing $\widetilde{\mathbf{W}}$.

This procedure is the same for the two-step and the direct approach, with the only distinction in the definition of $\tilde{\mathbf{x}}_u$ in (13). For the two-step approach this is simply the unique word in frequency domain, added in (3):

$$\tilde{\mathbf{x}}_u = \mathbf{F}_N \begin{bmatrix} \mathbf{0} \\ \mathbf{x}_u \end{bmatrix} \quad (15)$$

For the direct approach this derives from (11):

$$\tilde{\mathbf{x}}_u = \mathbf{F}_N \mathbf{B} \mathbf{P} \begin{bmatrix} \mathbf{0} \\ \mathbf{M}_{22}^{-1} \end{bmatrix} \mathbf{x}_u \quad (16)$$

The remaining procedure does not depend on the used UW generation method.

## III. SYMBOL ENERGIES IN UW-OFDM

Although the two presented approaches how to generate the unique word for the OFDM symbol seem almost identical, the impact on the symbol energies is tremendous, as we will show in this section, based on [3].

## A. Symbol energy for the two-step approach

Using the expectation operator $\mathrm{E}\{\cdot\}$, Parseval's theorem, which allows us to derive the symbol energy in frequency domain and omittint the matrices $\mathbf{BP}$, that do not change the energy budget, the mean energy of an OFDM symbol, when averaging over all possible data vectors, is given by

$$\begin{aligned} E_\mathbf{x} &= \mathrm{E}\left\{ \mathbf{x}^\mathsf{H} \mathbf{x} \right\} = \frac{1}{N} \mathrm{E}\left\{ \tilde{\mathbf{x}}^\mathsf{H} \tilde{\mathbf{x}} \right\} + \mathbf{x}_u^\mathsf{H} \mathbf{x}_u \\ &= \frac{1}{N} \mathrm{E}\left\{ \begin{bmatrix} \tilde{\mathbf{x}}_d^\mathsf{H} & \tilde{\mathbf{x}}_r^\mathsf{H} \end{bmatrix} \begin{bmatrix} \tilde{\mathbf{x}}_d \\ \tilde{\mathbf{x}}_r \end{bmatrix} \right\} + \mathbf{x}_u^\mathsf{H} \mathbf{x}_u \\ &= \underbrace{\frac{1}{N} \mathrm{E}\left\{ \tilde{\mathbf{x}}_d^\mathsf{H} \tilde{\mathbf{x}}_d \right\}}_{E_d} + \underbrace{\frac{1}{N} \mathrm{E}\left\{ \tilde{\mathbf{x}}_r^\mathsf{H} \tilde{\mathbf{x}}_r \right\}}_{E_r} + \underbrace{\mathbf{x}_u^\mathsf{H} \mathbf{x}_u}_{E_u}. \end{aligned} \quad (17)$$

The data symbols are assumed to be uncorrelated and from a QAM alphabet with zero mean and variance $\sigma_d^2$ which provides $E_d = \frac{N_d \sigma_d^2}{N}$.

The trace operation $\mathrm{tr}(\cdot)$, which is the sum of the main diagonal elements of a matrix, and (6) help us to rewrite the energy of the redundant carriers as

$$\begin{aligned} E_r &= \frac{1}{N} \mathrm{E}\left\{ \tilde{\mathbf{x}}_r^\mathsf{H} \tilde{\mathbf{x}}_r \right\} = \frac{1}{N} \mathrm{E}\left\{ \mathrm{tr}\left( \tilde{\mathbf{x}}_r \tilde{\mathbf{x}}_r^\mathsf{H} \right) \right\} \\ &= \frac{1}{N} \mathrm{tr}\left( \mathrm{E}\left\{ \tilde{\mathbf{x}}_r \tilde{\mathbf{x}}_r^\mathsf{H} \right\} \right) = \frac{1}{N} \mathrm{tr}\left( \mathrm{E}\left\{ \mathbf{T} \tilde{\mathbf{x}}_d \tilde{\mathbf{x}}_d^\mathsf{H} \mathbf{T}^\mathsf{H} \right\} \right) \\ &= \frac{1}{N} \mathrm{tr}\left( \mathbf{T} \mathrm{E}\left\{ \tilde{\mathbf{x}}_d \tilde{\mathbf{x}}_d^\mathsf{H} \right\} \mathbf{T}^\mathsf{H} \right) \\ &= \frac{\sigma_d^2}{N} \mathrm{tr}\left( \mathbf{T} \mathbf{T}^\mathsf{H} \right). \end{aligned} \quad (18)$$

The amount of energy needed for the redundant symbols depends on $\mathbf{T}$, which is only influenced by the number and positions of the redundant carriers.

## B. Symbol energy for the direct approach

For the direct approach we start the derivation in frequency domain:

$$\begin{aligned} E_\mathbf{x} &= \frac{1}{N} \mathrm{E}\left\{ \tilde{\mathbf{x}}^\mathsf{H} \tilde{\mathbf{x}} \right\} \\ &= \frac{1}{N} \mathrm{E}\left\{ \tilde{\mathbf{x}}_d^\mathsf{H} \tilde{\mathbf{x}}_d \right\} + \frac{1}{N} \mathrm{E}\left\{ \tilde{\mathbf{x}}_r^\mathsf{H} \tilde{\mathbf{x}}_r \right\} \\ &= \frac{N_d \sigma_d^2}{N} + \frac{1}{N} \mathrm{E}\left\{ \tilde{\mathbf{x}}_r^\mathsf{H} \tilde{\mathbf{x}}_r \right\}. \end{aligned} \quad (19)$$

We use now (10) to further get

$$\begin{aligned}
E\{\tilde{\mathbf{x}}_r^H \tilde{\mathbf{x}}_r\} &= E\{\text{tr}(\tilde{\mathbf{x}}_r \tilde{\mathbf{x}}_r^H)\} = \text{tr}(E\{\tilde{\mathbf{x}}_r \tilde{\mathbf{x}}_r^H\}) \\
&= \text{tr}\left(E\left\{(\mathbf{T}\tilde{\mathbf{x}}_d + \mathbf{M}_{22}^{-1}\mathbf{x}_u)(\mathbf{T}\tilde{\mathbf{x}}_d + \mathbf{M}_{22}^{-1}\mathbf{x}_u)^H\right\}\right) \\
&= \text{tr}\Big(E\Big\{\mathbf{T}\tilde{\mathbf{x}}_d \tilde{\mathbf{x}}_d^H \mathbf{T}^H + \mathbf{M}_{22}^{-1}\mathbf{x}_u \tilde{\mathbf{x}}_d^H \mathbf{T}^H \\
&\quad + \mathbf{T}\tilde{\mathbf{x}}_d \mathbf{x}_u^H (\mathbf{M}_{22}^H)^{-1} + \mathbf{M}_{22}^{-1}\mathbf{x}_u \mathbf{x}_u^H (\mathbf{M}_{22}^H)^{-1}\Big\}\Big) \\
&= \sigma_d^2 \text{tr}(\mathbf{T}\mathbf{T}^H) + \text{tr}(\mathbf{M}_{22}^{-1}\mathbf{x}_u \mathbf{x}_u^H (\mathbf{M}_{22}^H)^{-1}) \quad (20)
\end{aligned}$$

and finally obtain

$$E_{\mathbf{x}} = \underbrace{\frac{N_d \sigma_d^2}{N}}_{E_d} + \underbrace{\frac{\sigma_d^2}{N}\text{tr}(\mathbf{T}\mathbf{T}^H)}_{E_r} + \underbrace{\frac{1}{N}\mathbf{x}_u^H (\mathbf{M}_{22}^H)^{-1} \mathbf{M}_{22}^{-1} \mathbf{x}_u}_{E_u}. \quad (21)$$

We identify the same expressions for $E_d$ and $E_r$ as in the two step approach. The diference is in $E_u$ which is the only term that depends on the actual choice of the unique word.

In [3] it is shown that the inequality

$$\mathbf{x}_u^H \mathbf{x}_u \leq \frac{1}{N} \mathbf{x}_u^H \mathbf{M}_{22}^{-H} \mathbf{M}_{22}^{-1} \mathbf{x}_u \quad (22)$$

holds in any case. Hence the OFDM symbol from the two-step approach always needs equal or less energy than the symbol generated by the direct approach.

We emphasize the difference between the energy of the unique word $\mathbf{x}_u^H \mathbf{x}_u$ and the energy effected by the generation of the unique word $E_u$. The latter is contained in $\mathbf{x}_u$ and $\tilde{\mathbf{x}}_p$ in the time domain symbol, or in $\tilde{\mathbf{x}}_r$ only when looking at the frequency domain representation. We will call the difference between the left and the right hand side *excess energy*, which is only present in $\mathbf{x}_p$ or $\tilde{\mathbf{x}}_r$.

## IV. NUMERICAL EXAMPLES

With (22) it is obvious that the energy of an OFDM symbol generated by the two-step approach is always equal or lower than the energy of a symbol generated by the direct approach. We also assume the excess energy of the direct approach is wasted and does not contribute to transmission reliability in terms of bit error ratio (BER), since this energy is concentrated on the redundant carriers only.

In this section we want to give some insight to the dimension of this issue for practical setups. Therefore we consider three common unique word sequences, compare their energy consumption in the two-step and the direct approach and finally show their impact on the BER.

The BER curve of a CP-OFDM system according to IEEE 802.11a [4] will be included as a reference. We apply the same parameters for UW-OFDM wherever possible, i.e. $N = 64$, the length of the guard interval will be the unique word length $N_u = N_r = 16 = N_r$ and 12 zero carriers will be included at the band edges and the DC carrier. IEEE 802.11a also includes four pilot carriers with an overall energy of $4/52$ of the whole OFDM symbol. Since in UW-OFDM we aim to use the unique word for synchronization tasks, we scale the UW to the same percentage for comparison reasons.

The unique word sequences used for comparison in this work are:

1) The generalized Barker sequence [7] of length 12 padded with zeros to the final length of 16.
2) A CAZAC sequence (constant amplitude, zero autocorrelation) from [5], as often used for channel estimation, frequency offset estimation and timing synchronisation.
3) The length 16 unique word from [6], which also has CAZAC properties.

The average energy demand per OFDM symbol when using these sequences as unique words is shown in figure 2, split in data energy $E_d$, redundant energy $E_r$ and UW generation energy $E_u$.

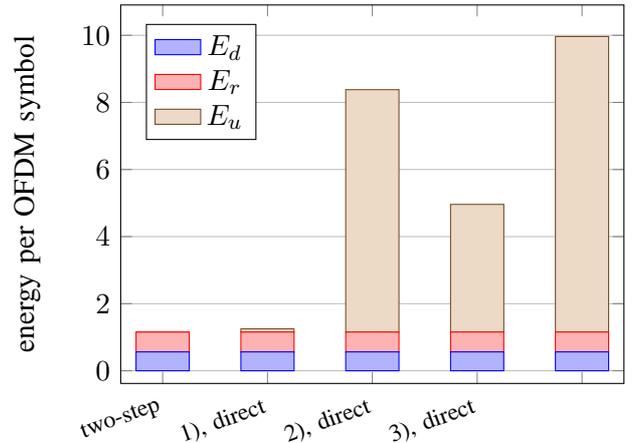

Fig. 2. Symbol energies for different unique words and approaches

The first bar shows the symbol energy if the zero word is used as UW. Since all UW energies $\mathbf{x}_u^H \mathbf{x}_u$ are normalized, the two-step approach yields the same symbol energy for any UW, according to (17). Thus the second bar represents this case with only a barely noteable $E_u$ bar, being in fact $4/52$ of the whole, topping the zero word energy. The three remaining bars show the average energies of the OFDM symbols generated by the direct approach for the chosen UWs. We note the huge amount of excess energy needed only by using the direct approach for

UW generation. As stated earlier, the excess energy depends on the particular UW design and on the positions of the redundant carriers.

Figure 3 shows the performance of an uncoded transmission over an AWGN channel. The bit error rate is plotted against $E_b/N_0$.

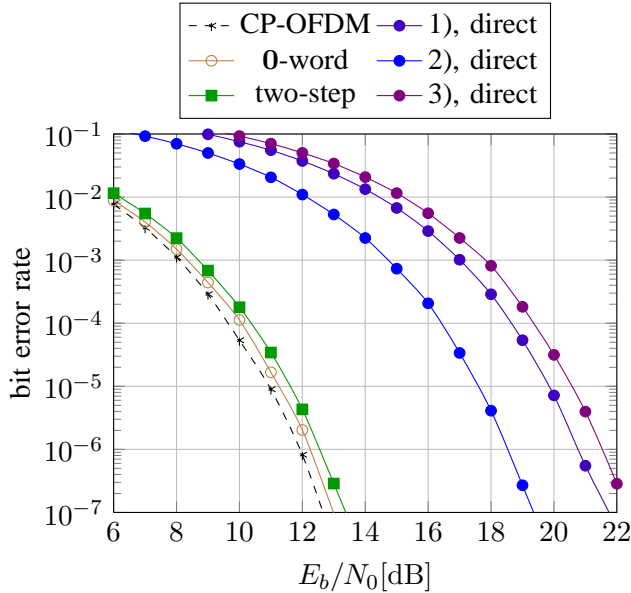

Fig. 3. BER performance using the different UWs and different symbol generation approaches

Since the symbols generated by the two-step approach have the same energies, the curves coincide and only one representative is shown. The bit error performances of the direct approach transmissions suffer from the huge excess energy. If we compare the curve of sequence No. 3), direct vs. two-step approach, we note a constant shift of about 9dB. This exactly coincides to the corresponding symbol energies

$$\frac{E_{\mathbf{x}}^{(3)}}{E_{\mathbf{x}}^{\text{(two-step)}}} = \frac{9.96}{1.25} = 7.97 \triangleq 9.01\text{dB}. \quad (23)$$

From figure 3 it can be seen that the simulated UW-OFDM system (using the two step approach) shows a small degradation over the reference CP-OFDM system (IEEE 802.11a) in an AWGN environment. However UW-OFDM shows its potential in frequency selective environments. Figure 4 shows results (taken from [3]) for two different indoor multipath scenarios, one featuring deep spectral fades (channel 1), and one representing a nearly flat fading channel (channel 2). Note that all curves are generated without using an outer channel code, furthermore the two step approach is applied in UW-OFDM. Especially for channel 1 UW-OFDM significantly outperforms CP-OFDM.

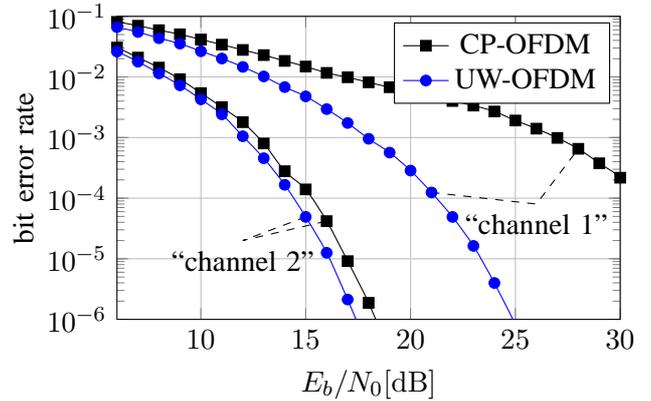

Fig. 4. BER performance comparing UW-OFDM with the reference CP-OFDM system in frequency selective scenarios

## V. CONCLUSION

In this work we had a closer look on two possibilities of unique word generation in UW-OFDM and derived analytical expressions for the corresponding OFDM symbol energies. With the aid of three particular example UW sequences, we showed that the symbol energies of OFDM symbols generated by the direct approach need much more energy than those generated by the two-step approach. The excess energy of the direct approach is completely wasted, since it does not improve the transmission reliability.

Thus we will exclude the direct approach in further research.